\documentclass[12pt,preprint]{aastex}
\usepackage{amsmath}
\usepackage{natbib}

\begin{document}

\bibliographystyle{apj}

\title{{\it MOST} Spacebased Photometry of the Transiting Exoplanet System
        HD 209458: Transit Timing to Search for Additional Planets}

\author{Eliza Miller-Ricci}

\affil{Harvard-Smithsonian Center for Astrophysics, 60 Garden St. Cambridge,
        MA 02138}

\email{emillerricci@cfa.harvard.edu}

\author{Jason F. Rowe}

\affil{University of British Columbia, 6224 Agricultural Road, Vancouver, BC
        V6T 1Z1, Canada}

\author{Dimitar Sasselov}

\affil{Harvard-Smithsonian Center for Astrophysics, 60 Garden St. Cambridge,
        MA 02138}

\author{Jaymie M. Matthews}

\affil{University of British Columbia, 6224 Agricultural Road, Vancouver, BC
        V6T 1Z1, Canada}

\author{David B. Guenther}

\affil{Department of Astronomy and Physics, St. Mary's University, Halifax, NS
        B3H 3C3, Canada}

\author{Rainer Kuschnig}

\affil{Department of Physics and Astronomy, University of British Columbia,
        6224 Agricultural Road, Vancouver, BC V6T 1Z1, Canada}

\author{Anthony F.J Moffat}

\affil{D\'{e}partement de Physique, Universit\'{e} de Montr\'{e}al, C.P. 6128,
       Succ. Centre-Ville, Montr\'{e}al, QC H3C 3J7, Canada}

\author{Slavek M. Rucinski}

\affil{David Dunlap Observatory, University of Toronto, P.O. Box 360,
       Richmond Hill, ON L4C 4Y6, Canada}

\author{Gordon A.H Walker}

\affil{Department of Physics and Astronomy, University of British Columbia,
        6224 Agricultural Road, Vancouver, BC V6T 1Z1, Canada}

\author{Werner W. Weiss}

\affil{Institut f\"{u}r Astronomie, Universit\"{a}t Wien,
       T\"{u}rkenschanzstrasse 17, A-1180 Wien, Austria}

\begin{abstract}

We report on the measurement of transit times for the HD 209458 planetary
system from photometry obtained with the {\it MOST} \footnote{Based on data 
from the {\it MOST} satellite, a Canadian Space Agency mission, jointly 
operated by Dynacon 
Inc., the University of Toronto Institute for Aerospace Studies and the 
University of British Columbia, with the assistance of the University of 
Vienna.} (Microvariability \& 
Oscillations of STars) space telescope. Deviations
from a constant orbital period can indicate the presence of additional
planets in the system that are yet undetected, potentially with masses
approaching an Earth mass.  The {\it MOST} data sets of HD 209458 from 
2004 and 2005 represent unprecedented time coverage with nearly 
continuous observations spanning 14 and 43 days and monitoring 3 transits and 
12 consecutive transits, respectively.  The transit times we obtain show no
variations on three scales: (a) no long-term change in P since before 2004
at 25 ms level, (b) no trend in transit timings during the 2005 run, and
(c) no individual transit timing deviations above 80 sec level.  Together with 
previously published transit times from \citet{ago07}, this allows us
to place limits on the presence of additional close-in
planets in the system, in some cases down to below an Earth mass.
This result, along with previous radial velocity work, now eliminates the 
possibility that a perturbing planet could be
responsible for the additional heat source needed to explain HD 209458b's
anomalous low density.

\end{abstract}


\section{Introduction}

The search for extrasolar planets has been characterized recently
by a move toward the discovery of lower mass planets compared to the gas
giant planets of our Solar System.  While the techniques of
radial velocities, transits, and microlensing have already
proved to be successful in detecting extrasolar planets, the novel
idea of using transit timing variations (TTVs) to detect additional low-mass
companions in known transiting planetary systems \citep{ago05, hol05}
is a recent and exciting addition to the repertoire of planet detection 
methods, which has not yet been widely applied.  The premise of the TTV method 
is that a planet known to transit its host star will do so at a near-constant 
period unless there is an additional planet in the system to perturb its 
orbit.  In the case of a second planet in the system, the observer
will then view transits that deviate from a constant period on a level of
seconds to minutes, depending on the mass and orbital parameters of the 
perturbing planet.   An analysis of transit timings in the TrES-1 system was
carried out by \citet{ste05} placing useful limits on the presence of
additional planets around that star.  They determine that a (2:1 resonant) 
companion planet
in that system would generally need to have a mass comparable to or less than
an Earth mass.  A subsequent analysis performed on HST observations of HD 
209458 by \citet{ago07} places similar limits on the presence of companion 
planets in that system.  The addition of the {\it MOST} data now allows for 
more stringent limits to be placed on such planets.

HD 209458 was the first star found to have a transiting planet
\citep{cha00, hen00} and remains the second brightest star known to have a
transiting planetary companion, making it suitable for a large number of 
follow-up observations.  As more
transiting planets are discovered, it is becoming clear that HD 209458b
has an anomalously large radius for its mass when compared
with theoretical predictions \citep{lau05c}.  Speculation as to the
mechanism that is keeping the planet ``too large'' has led to several theories
including (i) additional sources of internal heating
such as strong winds \citep{gui02}, (ii) obliquity tides due to a disalignment
of the planetary spin axis and the orbit normal \citep{win05}, and
(iii) the presence of additional low-mass planets in the system inducing
eccentricity in the orbit of HD 209458b and causing ongoing tidal heating
\citep{bod01, bod03}.  The
first of these theories seems less likely, seeing as it would apply
to the radii of all of the transiting planets.  The latter two have remained
viable theories in explaining the anomalous radius of HD 209458b.  

The TTV analysis that follows serves to shed some light on the presence of
additional planets in the system, and, together with previous observations,
now effectively rules out the third option listed above.  The Microvariability 
and Oscillations of STars ({\it MOST}) satellite \citep{wal03, mat04} observed
HD 209458 in 2004 and 2005 as it passed through the satellite's Continuous
Viewing Zone (CVZ).  A combined total of 15 complete transits were
observed, with 12 of these being consecutive transits in 2005.  The combination
of the timings for these transits along with previously published transit 
times by \citet{ago07} allow us to place strong limits on the presence of
additional planets in the system, which would be undetectable by other 
currently available techniques.

\section{{\it MOST} Photometry}

Photometry of HD 209458 was obtained with {\it MOST} during 13.5 days in 2004 
and 42.9 days in 2005.  {\it MOST} houses a 15-cm optical telescope feeding a 
CCD photometer through a single custom broadband optical filter.  From its 
820-km Sun-synchronous polar orbit, it can monitor stars passing through its 
CVZ for up to two months without interruption.  {\it MOST} can collect 
photometry in several
operating modes, including ``Direct Imaging'', in which a target star is
centred in a subraster on the {\it MOST} Science CCD for a combination of 
aperture and PSF (Point Spread Function) photometry.

HD 209458 was observed in Direct Imaging mode, with exposures of 1.5 s
(as outlined in \citet{row06}).  The point-to-point precision for these
observations depends on the level of stray light scattered into the instrument 
and can be as low as 3 mmag at times of very little stray light and as high as 
20 mmag during instances of high stray light.  At low levels of stray light
this corresponds to a photometric error that is essentially equal to the
Poisson noise limit, whereas during times of increased scattered light this 
can degrade to up to 1.5 times the Poisson limit (see Figure 2 in 
\citet{row07}).  For the 2004 observations the sampling rate was varied to 
obtain a higher
rate during phases of the known planet's secondary eclipse, resulting in an
average sampling of 5 individual exposures per minute.  For the 2005 
photometry, the sampling rate
was 6 times per minute, a limit set by the data downlink capacity of the 
{\it MOST} satellite in the Direct Imaging mode.

The reduction procedures for the 2004 photometry are described by \citet{row06}
and include corrections for stray light  due to Earthshine modulated with the
satellite orbit.  In the 2005 data, there was electronic 
``crosstalk'' between the timing
of the Science CCD and the Attitude Control System (ACS) CCD onboard
electronics, with a timescale of about 3 days.  This manifested itself as a
subtle band of noise moving across the CCD subraster in about half a day.
Such
crosstalk added a non-Poisson component to the photometric noise, and
uncorrected could induce deviations of about 1 mmag on timescales of up to
a day.
By comparing the standard deviation of the background sky counts in the
subraster
against the square root of the mean sky count, we can trace the occurance of
crosstalk.  Corrections are then applied to the sky counts and the
instrumental magnitudes of HD 209458.  Other than this,
the 2004 and 2005 datasets have been uniformly reduced.

\begin{figure}
\plotone{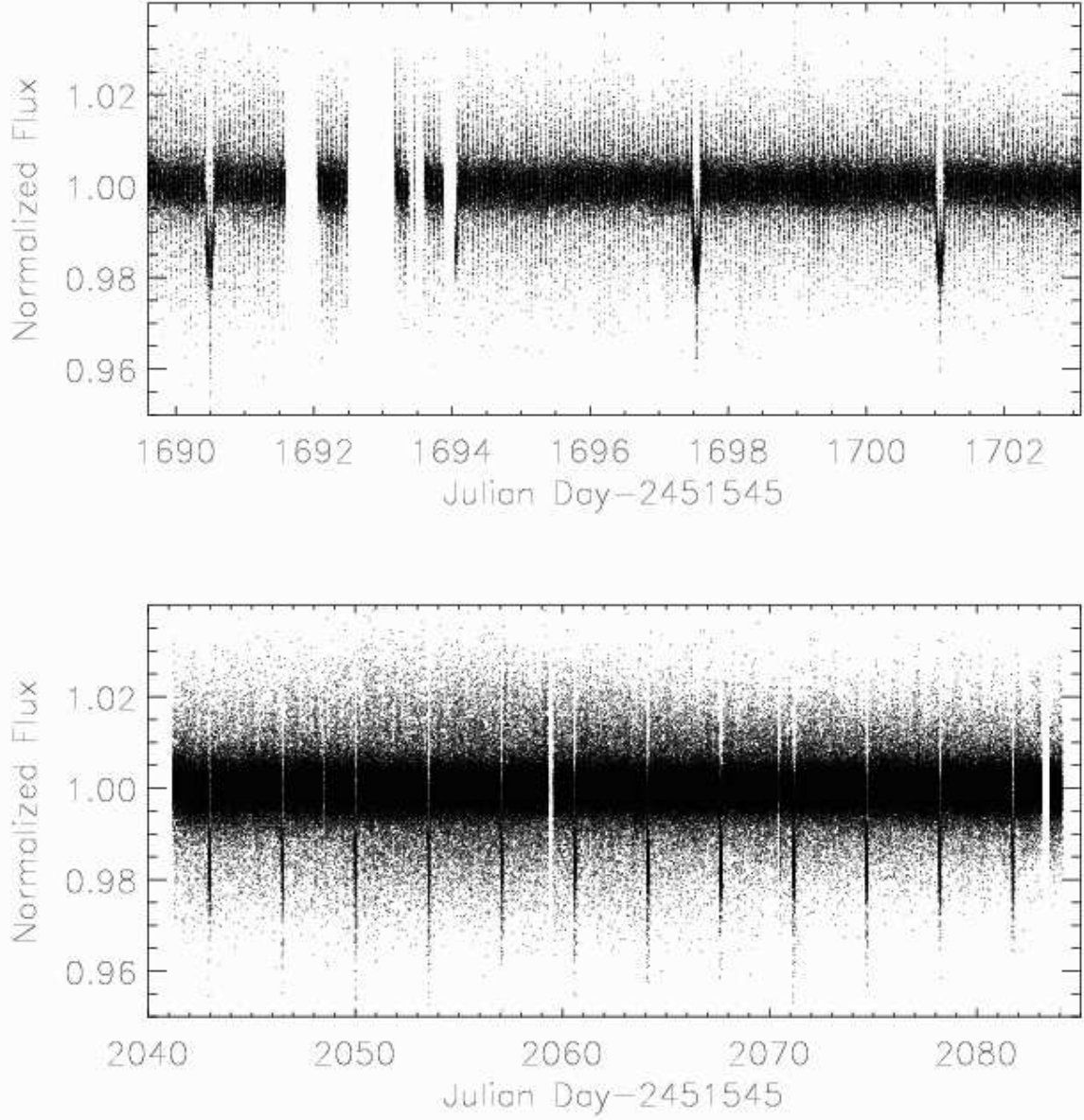}
\caption{Unbinned {\it MOST} HD 209458 photometry from 2004 (top) and 2005
  (bottom).  Note the difference in time axis scales between the two panels.  
  Incidences of very high stray light and  passages through
  the SAA, have been removed.  Gaps in the lightcurve during 2004 are due to
  onboard crashes during new software testing, while the two shorter gaps in 
  2005 were removed due to particularly high cosmic ray fluxes.
        \label{lightcurve}}
\end{figure}

For this analysis, we have removed from the data set incidences of very high
levels of stray light (Earthshine) modulated at the satellite's orbital
period,
as well as times of passage through the South Atlantic Anomaly (SAA) when the
cosmic ray flux is high, and other instances of severe cosmic ray hits. This
reduces the total number of observations by about 12\%, but does not
seriously
reduce the coverage of the observations over the HD 209458b orbital period.
The reduced 2004 and 2005 light curves are shown in Figure~\ref{lightcurve}.

The pronounced gaps seen in Figure~\ref{lightcurve} in the 2004 data are
due to
onboard crashes caused by the testing of new software during those trial
observations.  The two shorter noticeable gaps in the 2005 light curve are
due
to especially severe SAA passages, which were removed from the data.  There is
some evolution of the level of scattered light during the 2005 data, dropping
slightly toward the end of the run, consistent with seasonal evolution
over six
weeks that has been observed for other targets.  There is also an unfortunate
coincidence for {\it MOST} observations of HD 209458, in that the giant planet
orbits at a period that is almost exactly 50 times the 101.413 min 
orbital period of {\it MOST}.  Hence,
when we remove portions of the light curve with high stray light, these
consistently lie at the same phases in the light curve phased to the planet
transit period. We are alert to the possible effects on the transit timing
analysis, and have found that our results are not very sensitive to this
effect (see below).

\section{Transit Times for HD 209458b}

\begin{figure}
\plotone{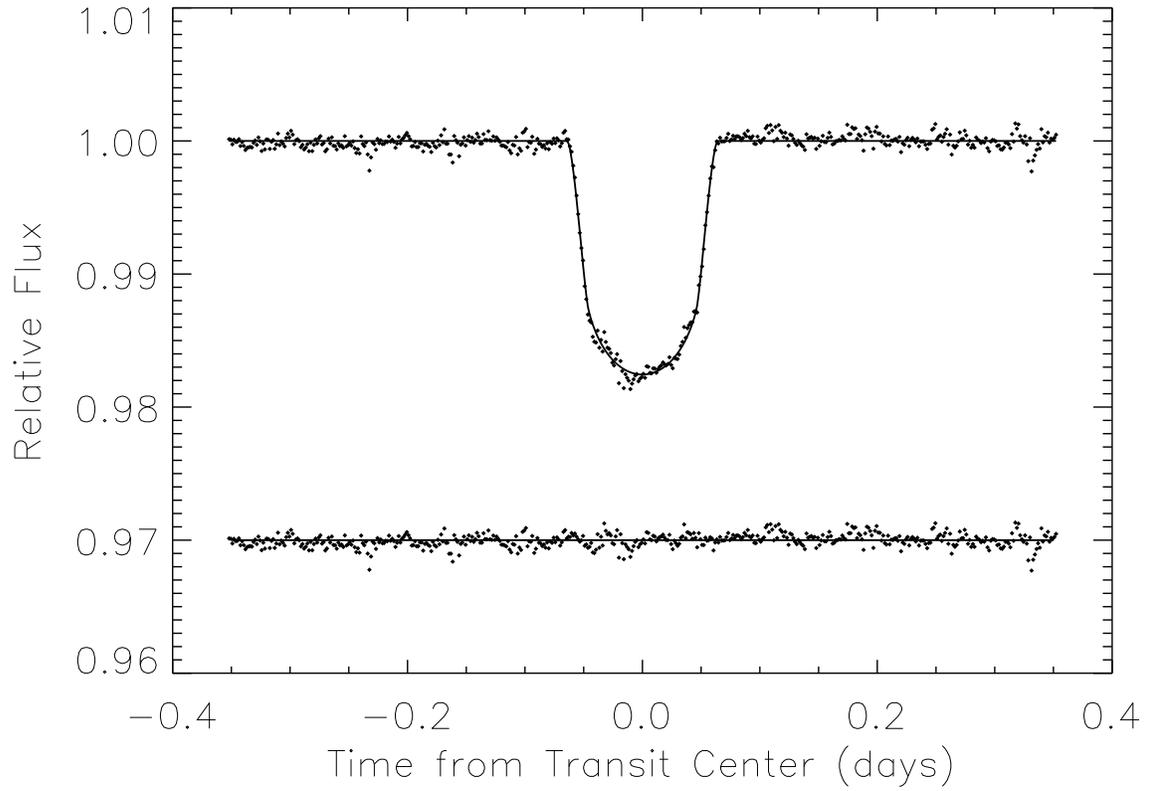}
\caption{The combined 2004 and 2005 data overlaid with the transit model
        (above) and residuals from this model (below).  The data has
        been phased at the orbital period of the planet and binned at a 2
        min interval.
        \label{Model}}
\end{figure}

In computing our transit times we compare the {\it MOST} data against a model
transit curve, constructed using the formalism set forth in
\citet{man02} for a source with nonlinear limb darkening.  To determine the
orbital parameters for the model we start with those
laid out in \citet{knu07}, which in turn were determined by a multi-parameter
fit to data from the STIS
spectrometer aboard the Hubble Space Telescope (HST).  Noting that the
{\it MOST} transit light curve appears to be slightly deeper than this model 
would predict, we then fit for the stellar and planetary radii by minimizing 
the chi-square statistic for both variables on the composite phased and binned 
{\it MOST} light curve.  Our values for the two radii differ somewhat
from the HST light curve but lie within the 1-$\sigma$ error
bars determined for that data set, with the {\it MOST} data implying a 
planetary radius of 1.339 R$_J$ and a stellar radius of 1.118 R$_{\sun}$ 
(see discussion in \citet{row07}).  Nonlinear limb darkening parameters for 
the {\it MOST} bandpass are derived from synthetic spectra calculated by R. 
Kurucz
\footnote{See http://kurucz.harvard.edu/stars/hd209458}.  This is accomplished
by summing the spectra across the disk of the star then multiplying by the 
throughput of the {\it MOST} optics and detector.
The resultant transit model has system
parameters that are given in Table~\ref{tbl1}, and this model is overlaid
on the phased and binned data in Figure~\ref{Model}.  We note that some
correlation in the residuals from this model (red noise) remains.   This is
most likely due to the stray light experienced by {\it MOST},
as it repeats at approximately the orbital period of the satellite, despite 
reduction procedures to correct for this effect.  We pay particularly close 
attention to the modulation of stray light to determine its effects on the 
transit timing measurements as decribed below.  

\begin{figure}
\plotone{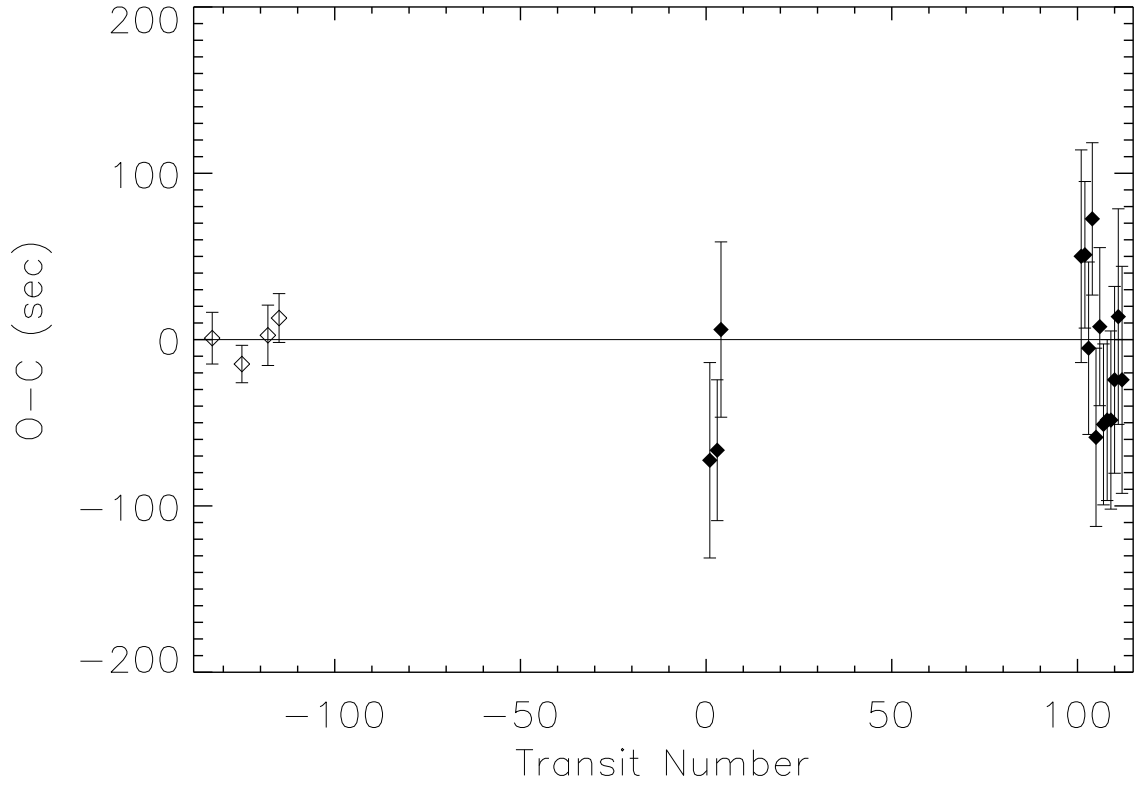}
\caption{Deviation from predicted time of transit vs. transit number for
        the 2004 and 2005 {\it MOST} data (filled symbols) and for the 2003 
	HST data as reported by \citet{ago07} (open symbols).  The expected 
	time of transit is based on the ephemeris of \citet{knu07}.  
        \label{O-C}}
\end{figure}

\begin{figure}
\plotone{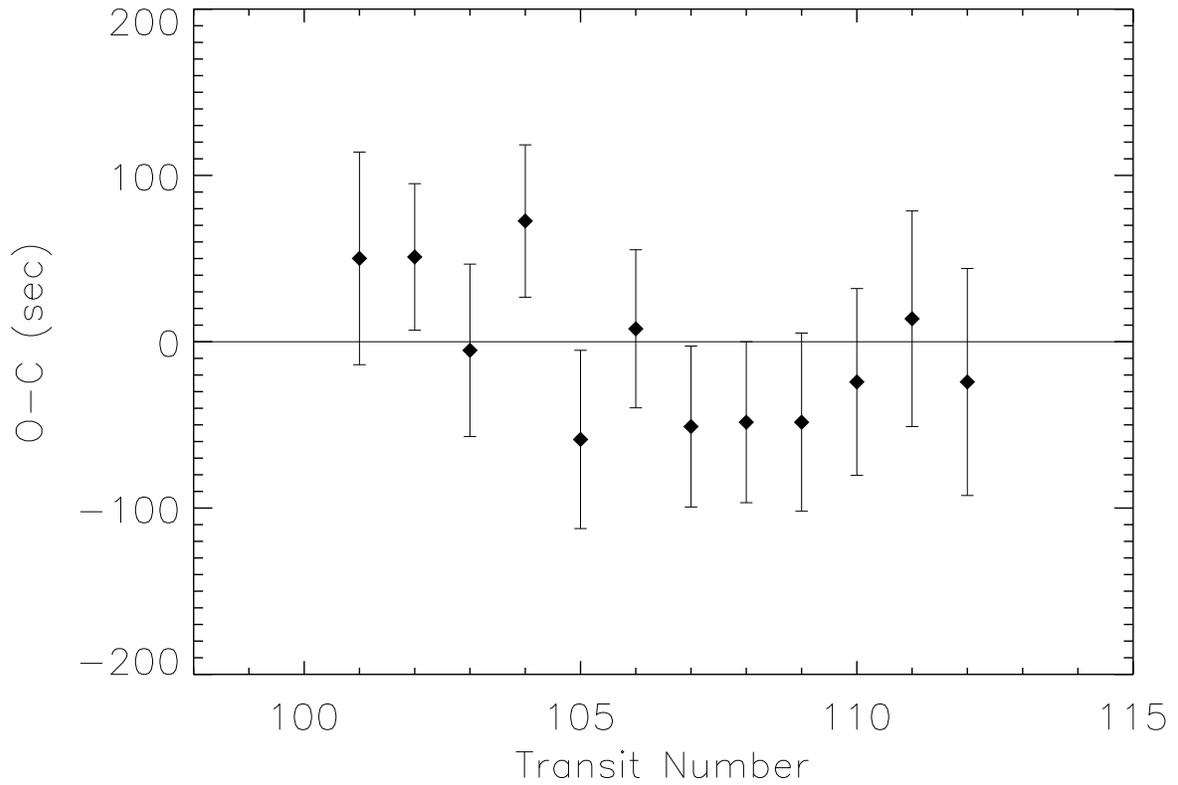}
\caption{Deviation from predicted time of transit vs. transit number for the
        12 transits in the 2005 data only.  The short-term scatter
	in the transit times here is always less than $100 \, {\rm s}$ and is 
	less than 30 s for the last 7 transits where scattered light entering
	the instrument is at a minimum.
        \label{O-C_2005}}
\end{figure}

To determine times for each of the individual transits, the model light curve 
is computed at impact parameters
corresponding with each of the {\it MOST} data points.  We then find the
center-of-transit time for each transit at which the $\chi^2$ value for the
fit to the data is minimized.  The resultant times
are presented in Table~\ref{tbl2} in HJD, 
calculated using the IDL function helio\_jd \citep{lan93}.  
We measured transit times for each of the 3 complete transits in the 2004
data and the 12 transits in 2005, then calculated the timing
difference from the expected times of transit, as shown in Figures~\ref{O-C}
and~\ref{O-C_2005}.  The expected times of transit are calculated by
adding the correct number of orbital periods of HD 209458b to the 2003
ephemeris from \citet{knu07}.  We note that our data agrees
with the orbital period and ephemeris of this previous work, since
our data points in Figures~\ref{O-C} and~\ref{O-C_2005} lie almost equally
above and below the line representing the expected times ($O - C = 0$).
As a reference, in Figure~\ref{O-C} we also show the times for the 4 
transits observed in 2003 with STIS as reported by \citet{ago07}, who used a
similar method to determine transit times.

Our error bars are calculated using a bootstrapping Monte Carlo simulation
similar to the one described in \citet{ago07}.  For each transit we 
shift the residuals from the best-fit transit model  (and their associated 
errors) by a random number of points.  We then add the new residuals back 
onto the transit model and recalculate the center-of-transit time using the 
same procedure described above, thus maintaining the original point-to-point 
correlations.  The goal realized by 
this error analysis is to understand the effects of correlated noise in the 
{\it MOST} light curve due to the repeating pattern of stray light.  The 
resulting error bars in Figures~\ref{O-C} and~\ref{O-C_2005} are somewhat 
larger than those obtained from a simple chi-squared analysis
(generally by less than a factor of 2, but occasionally by a factor of 
several).  In addition, we have performed tests of 
our sensitivity to the modulation of stray light during the {\it MOST} 
satellite orbit by applying
more stringent removal of times of higher stray light.  This does not have a
significant effect on the transit times we determine and only reduces
unnecessarily the overall duty cycle of the observations.  It must also be 
noted that, at the precision and time coverage of the {\it MOST} 
data, intrinsic low-amplitude variations in the star HD 209458a may be present.
This may represent a fundamental limit in the systematic errors of the timing
data for this system.

Long-term and periodic trends in the data not associated with the HD 209458b
orbit could be a serious problem in limiting the accuracy of our transit times
so we have examined this issue closely.  This is especially true if a deviation
due to instrumental or satellite orbital effects was to occur during transit
ingress or egress.  As an example, consider a transit lightcurve in
which ingress has not been correctly normalized so that all of the points
are reported at a flux level that is slightly too low.  In this case, our
analysis method would report this transit as occurring earlier than it
actually did.  We have carefully identified
and corrected for possible long-term and periodic artifacts in the data to
produce a properly normalized light curve,
as described in \citet{row06}.  These include
(i) dips and rises in the lightcurve stemming from electronic crosstalk as
described in Section 2; (ii) cycle-per-day modulation of stray Earthshine due 
to the Sun-synchronous orbit of {\it MOST}; and (iii) filtering of longer-term 
trends in the data which might be instrumental, stellar or a combination of 
both.  

Another consideration in determining transit times is the importance of
having a light curve that
is well-sampled during ingress and egress, as this is where the timing
signal is most sensitive.  In the
case of a partial transit that is missing either ingress or egress, the
value adopted for orbital inclination (which determines the transit
duration) in the model light curve must be finely tuned to the true value.
Otherwise, the risk is that partially sampled transits could have large 
systematic errors in their measured timings.  For the transit times 
reported from the {\it MOST} data only complete transits were used.  In 
particular, the second transit from the 2004 observations is not used in this
analysis since data are only available for half of it.  Note that the effect 
of removing short intervals of very high stray light, (which often affects 
ingress more than egress or vice versa) is negligible, since only a small 
number of data points are actually removed from each transit.

\section{Limits on Other Close-in Planets in the HD 209458 System}

The two seasons of transit timing data from {\it MOST} allow us to look for
variations due to several effects: (1) orbital decay of planet HD 209458b
\citep{sas03}; (2) precession of its orbit \citep{mir02, lau05b, hey07};
(3) moons \citep{sar99, bro01};  and (4) orbit perturbations caused by
additional planets in the system \citep{ago05, hol05, ste06}.  The first two
effects would cause long-term variations in the observed time between 
successive transits.  To check for these types of long-term drifts, our 
best-fit times for the 15 observed transits from
{\it MOST} can be combined with the 13 transit times reported by \citet{ago07}
to determine a revised orbial period for HD 209458b.  Fitting an orbital
period to these 28 transit times spanning from April 2000 to August 2005 yields
a period of 3.52474832 $\pm$ 0.00000029 days.  This is in
agreement with the 2003 orbital period reported by \citet{knu07} to within 
25 ms (within their 1-$\sigma$ error bars of 33 ms).  The expected drift in 
the orbital 
period of HD 209458b due to the first two effects in the list above over the 
two-year baseline obtained when comparing our orbital period to 
the 2003 period of \citet{knu07} remains smaller than the current 
sensitivity. 

The latter two effects on the list presented above will cause short-term
timing variations.  In terms of effect (4), we set our approach to analyzing 
HD 209458 as follows.
Current Doppler radial velocity (RV) data already exclude planets with
$M > 3\times 10^{-4}M_{\odot}$ ($100$ M$_{\earth}$) and $P < 100$ days 
\citep{lau05}, and at such long-period orbits RVs are
more sensitive than TTVs. This interplay between RV and TTV limits was pointed
out by Agol et al.(2005) - their Fig.7, and is likely to be common to other
bright transiting systems with good RV data (e.g., HD 189733, TrES-1, HAT-P-1).
Therefore the set of {\it MOST} transit times is particularly fit to place
limits on small close-in perturbing planets, with $P < 15$ days (or about 
a third
of the length of the observing run). Despite its limitations, this is of great
interest for the HD 209458 system, because such small close-in planets could
still provide enough forcing to the eccentricity of HD 209458b for tidal
heating to occur. The same is true for systems like HAT-P-1b.


Our results show no evidence for short-term timing variations.  The 
short-term scatter in the data is always below 100 s and is below 30 s for the 
last 7 transits in 2005.
This imposes strong constraints on any additional planets - both in mass and
orbits.  To determine what types of planets are ruled out by the {\it MOST} 
data we solved the classical N-body problem,
\begin{equation}
\frac{d^2x_i}{dt^2} = - \Sigma_{j=1;j\neq i}{N} \frac{Gm_j (x_i -
x_j)}{|x_i-x_j|^3},
\end{equation}
where for 3 bodies, x describes the initial positions of the
particles.  For the HD 209458 system, we assumed a stellar mass of 1.101
$M_\sun$ and planetary mass of 0.69 $M_{jup}$ in a circular orbit with a
period of 3.52474832 days.  A third body was inserted with an initially
circular orbit, with periods ranging from 1 to 18 days in increments of
0.01 days and masses from 1 - 100 $M_\earth$ in 1 $M_\earth$ increments, and 
on coplanar orbits to HD 209458b.  
The solution was advanced at 1.0 second intervals for 1000 orbits of HD
209458b ($\sim 3\times 10^{8}$ s) using the LSODA routine from ODEPACK
\citep{rad93}.  In the case where the two planets in the simulation passed 
within 1 R$_{Jup}$ of each other, we determined that a collision had taken
place, and the N-body code was terminated. These orbits were deemed unstable
over the long term.  Resultant O-C values from the
simulations were calculated by a
linear interpolation to estimate the integration time when 
HD 209458b returns to the midpoint of crossing the disk of the star.  Also, in 
certain resonant orbits where we found that our data could limit the masses of
perturbing planets to less than an Earth mass, we ran a small number of 
additional N-body simulations down to 0.1 $M_\earth$ at 0.1 $M_\earth$
intervals to determine what ranges of sub-Earth mass planets were ruled out. 

\begin{figure}
\plotone{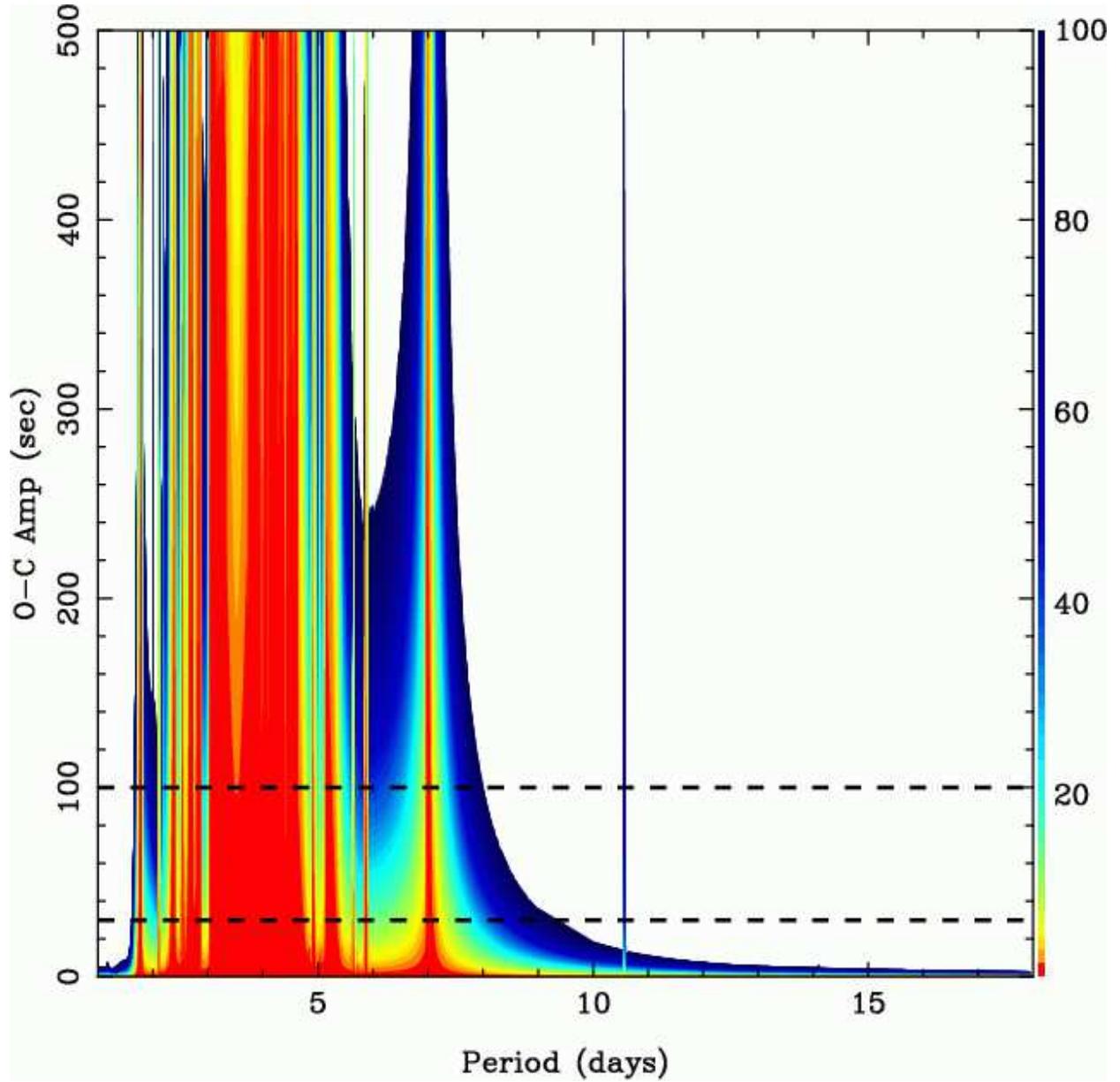}
\caption{N-body results for maximum transit timing deviation vs.\ orbital 
        period of the perturbing planet.  The color scale as defined on the
	right side of the plot indicates the mass of the perturbing planet
	(in M$_{\earth}$).  The 15 {\it MOST} transits show no timing
	deviations above the level of 100 s as indicated by the top dashed 
	line.  The last 7 transits observed are consistent with a constant 
	orbital period at the level of 30 s also indicated by a dashed line.  
        \label{JR_figure}}
\end{figure}

To determine the expected magnitude of the TTVs, we compute a Fourier 
Transform of the O-C series for each value
of period and mass of the 3rd body and extract the largest amplitude.  
This is shown in Figure~\ref{JR_figure}, where we plot the TTV amplitude vs.\ 
orbital period, resulting from the N-body simulations.  
Additionally, we compare the N-body results against all of the available 
transit timing data to determine the maximum mass that an additional planet 
in the 
HD 209458 sytem could possess, while still remaining consistent with the data.
This allows for robust mass limits to be placed on perturbing bodies
in the HD 209458 system. We fit the combination of the 15 {\it MOST} transit 
times and the 13 HST transit times \citep{ago07} to each of the O-C series 
generated by the 
N-body code.  This allows for a more complete coverage of the libration
period of any hypothetical 2-planet system, than fitting the {\it MOST} data 
alone.  We leave a timing offset and slope as free parameters.  From this
process, we determine the maximum mass of perturbing planet that remains 
consistent with the 1-sigma error bars on the transit times as a function
of the orbital period of the perurbing body, which we show in 
Figure~\ref{EMR_figure}. 

From Figure~\ref{EMR_figure}, we place the following constraints on the 
presence of additional planets in the HD 209458 system, at the 99\% (3-sigma) 
confidence limit.
First, we explore possible planets in inner orbits. Placed between
HD 209458b (at 0.045~AU) and the star, such planets are constrained to a very
narrow range of orbits. Non-zero eccentricity is a given but is also limited
for stability reasons. These orbits may be more likely due to resonant
trapping during migration \citep{zho05}, most likely at 1:2 with 
P$_{orb}\approx 1.76$ d . The transit timing data limits the mass of such 
planets to sub-Earth masses, $M > 0.9 \times 10^{-6}$ M$_{\sun}$ 
(about $0.3$ M$_{\earth}$).
If between the resonances, a planet could be as massive as 20 M$_{\earth}$ 
($6 \times 10^{-5}$ M$_{\sun}$) and escape detection in the transit timing 
data.

Previously undetected planets in outer orbits have a larger range of possible 
orbits and relaxed stability requirements on their eccentricity, compared to 
the inner ones discussed above.
Nice illustrations can be found in the detailed analysis of TrES-1 by 
\citet{ste05} and in Fig.\ 5 by \citet{ago05}. The {\it MOST} and HST 
transit data exclude sub-Earth mass planets down to 0.3 M$_{\earth}$ in the 
outer 3:2 and 2:1 mean-motion resonances,
and in the 3:1 resonance with $M > 8 \times 10^{-5}$ M$_{\sun}$
(40 M$_{\earth}$).  Outside of resonances, the {\it MOST} data limit any 
planets with $P < 7$ d to $M < 17$ M$_{\earth}$ and with 
$7 < P < 10$ d to $M < 100$ M$_{\earth}$. The exact limits depend on the period
of the perturbing planet according to Figure~\ref{EMR_figure}.  For orbits
with periods longer than 10 d the Doppler RV limits are better.

A range of intermediate orbits surrounding the 1:1 resonance with the 
transiting planet HD 209458b
are unstable for the lowest-mass perturbers, resulting in ejections or
collisions.  Between 2.65 and 4.65-day periods we can rule out sub-Earth
mass planets due to a combination of both TTV constraints and stability 
requirements.  

In addition to the N-body simulations described above, we performed
a limited number of additional calculations for perturbing planets in
either mutually inclined orbits relative to the transiting 
planet, or in initially eccentric orbits. For the case of a perturbing planet
on an initially circular orbit, we find that only small eccentricities
are attained (generally less than 0.1).  However, if even a small initial
eccentricity or mutual inclination between the planets is added, the resulting
orbit of the perturbing planet can attain a significantly higher mean 
eccentricity over the course of 1000 orbits.  The TTV's from these 
configurations therefore tend to be larger than those from the case presented 
above, of a perturbing planet on a circular orbit, coplanar with HD 209458b.  
Additionally, there is a smaller range of stable orbits in these cases.  
There are some limited cases
however where an eccentric companion can cause slightly larger TTV's than
a planet on a circular orbit.  As another consideration, perturbing planets 
starting from
different mean longitudes can result in TTV's that vary somewhat in amplitude,
where we always begin our N-body simulations with the second planet in an 
orbit 90 degrees out of phase with HD 209458b.
In the absence of these exceptions, the limits that we have placed on 
additional planets in the 
HD 209458 system are generally robust limits across the entire range of 
eccentricity parameter space, due to the fact that additional planets 
residing in eccentric or inclined orbits tend to have even larger observable 
effects on the transit times of HD 209458b.

\section{Summary and Conclusions}

\begin{figure}
\plotone{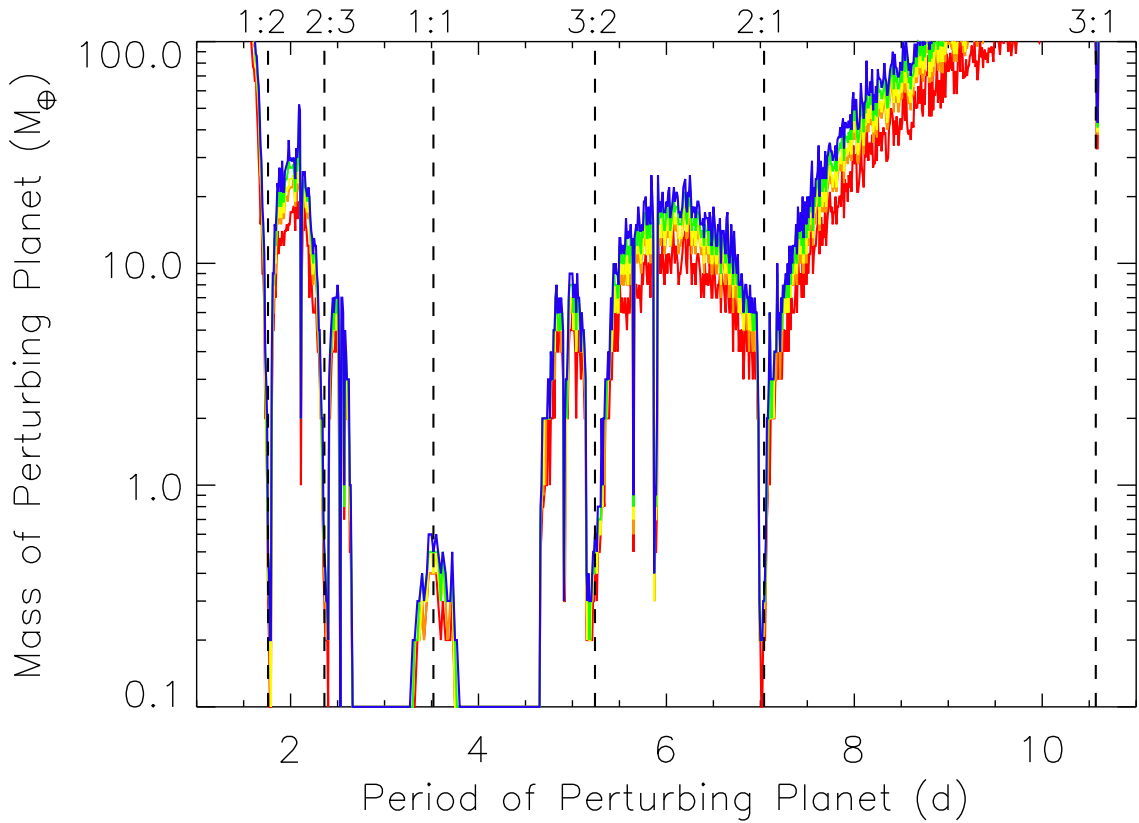}
\caption{Maximum mass allowed for a perturbing planet in the HD 209458 
        system, which still remains consistent with the {\it MOST} and HST
	transit times (from this work and \citet{ago07} respectively).  The 
	curves correspond to the masses of additional planets that are ruled 
	out at the 60\% (red), 90\% (orange), 95\% (yellow), 99\% (green), and 
	99.9\% (blue) confidence levels, based on the 1-$\sigma$ 
	error bars on the transit timing measurements.  Planets occupying the
	region of parameter space above the curves are ruled out by the 
	available transit timing data.
        \label{EMR_figure}}
\end{figure}

The addition of the 15 {\it MOST} transit times to the 13 previously 
available HST transit times from \citep{ago07} has allowed us to place
the tightest available limits on the presence of additional planets in
the HD 209458 system with orbital periods under 11 days.  We rule out
sub-Earth-mass planets in the inner 1:2 and 2:3 
resonances as well as in the outer 2:1 and 3:2 resonances.  Super-Earths are
excluded in a range of intermediate orbits as shown in Figure~\ref{EMR_figure}.

The {\it MOST} transit timing data are unique at this time in that they have 12
consecutive transits of uniform quality and consistent photometric
precision, which is very difficult to achieve from the ground for such long
time spans. For HD 209458 these types of observations have only been
possible from space and they have not been practical with the Hubble Space
Telescope, which can only observe partial transits.  Future space-based
missions designed to look at extrasolar planets, such as CoRoT and Kepler,
will also have the ability to detect additional low-mass planets in transiting 
systems via transit timing analysis. For the time being {\it MOST} serves as a
demonstration of the types of results that can be obtained from these types of 
observations.

Our results show no transit timing variations on three scales: (a) no long-
term change in P since 2003 at the 25 ms level, (b) no trend in transit timings
during the 2005 run, and (c) no individual transit timing deviations above the
80 sec level. No variations on scales (b) and (c) help exclude the presence of 
sub-Earth mass planets in inner resonant orbits and Earths and Super-Earths in 
outer resonances and close-in ($P<10$ d) orbits.  Therefore our results 
complement 
previous searches for a putative longer-period planet, e.g.~the 84-day period 
0.127$M_{Jup}$ perturber proposed by Bodenheimer et al.(2003). Our TTVs have 
no sensitivity to detect a planet like that, but the radial velocity data has 
already ruled this out \citep{lau05}.  Through a combination of the 15 
{\it MOST} transit times from 2004 and 2005 along with the 13 previously 
published transit times 
for HD 209458b obtained in 2000 to 2003 \citet{ago07}, we have
placed the most stringent limits on additional planets in this system with
orbits between 1.5 and 10 days.     

The level of orbital eccentricity of HD 209458b required to 
explain its anomalously large radius ranges between 0.012 and 0.03, 
given the uncertainty range for Jupiter's tidal dissipation factor, $Q_{J}$.
In order to induce such orbital eccentricity, a companion planet would either
need to be (i) significantly larger than Earth mass and in an outer orbit,
or (ii) at least Earth mass and in a mean-motion resonance with the transiting 
planet.  The first of these two cases is ruled out by radial velocity data 
\citep{lau05}.  The second case is ruled out by the TTV data. Hence the 
problem of the radius of HD 209458b still remains, with the presence of a 
sufficiently perturbing companion planet in the system ruled out and obliquity 
tides remaining as an attractive alternative explanation.

Our conclusions supporting the lack of additional planets in this system 
are further corroborated by the work of \citet{cro07} who use 
MOST photometry of the HD 209458 system to rule out 
the presence of additional {\it transiting} planets
with radii of 2-3 times that of the Earth and with periods ranging from 
0.5 days to two weeks.  While such planets would need to have suitable 
inclination angles to even be observed, the merit of a dual-transiting
system is that the combination of the radii of the planets (given by the
transit light curve) along with their masses (determined by their mutual 
interactions as seen in the TTVs) allows for the density of both planets in the
system to be determined unequivocally \citep{hol05}.  This application of
the transit timing method could potentially allow for the first determination
of the mass and density of an extrasolar Earth-like planet, since current RV
precision cannot detect such planets.

\acknowledgements

We would like to thank Heather Knutson and Matt Holman for valuable input
into determining and interpreting transit timing variations. We would also
like to thank our referee, Eric Agol, for his very useful comments on this 
paper.  JMM, DBG, AFJM,
SMR, and GAHW would like to acknowledge funding from the Natural Sciences \&
Engineering Research Council (NSERC) Canada.  RK is supported by the Canadian 
Space Agency.  WWW would like to thank the Austrian FFG (project MOST) and 
FWF (P17580) for funding.

\bibliography{ms}

\clearpage

\begin{deluxetable}{lc}
\tablecaption{Orbital and Physical Parameters for HD 209458 \label{tbl1}}
\tablewidth{0pt}
\tablehead{
\colhead{Parameter} & \colhead{Value}}
\startdata
M$_{*}$ [M$_{\sun}$] &  1.101 $\pm$ 0.064\tablenotemark{a}\\
R$_{*}$ [R$_{\sun}$] & 1.118 $\pm$ 0.03\\
R$_{pl}$ [R$_{Jup}$] & 1.339 $\pm$ 0.04\\
i [$^o$] & 86.929 $\pm$ 0.01\tablenotemark{a}\\
P [days] & 3.52474832 $\pm$ 0.00000029\\
c$_{1}$ & 0.410769 \\
c$_{2}$ & -0.108909 \\
c$_{3}$ & 0.904020 \\
c$_{4}$ & -0.437364 \\

\enddata

\tablenotetext{a}{From \citet{knu07}}
\tablecomments{The non-linear limb-darkening coefficients calculated
  for the {\it MOST} bandpass are given by c$_{1}$-c$_{4}$.}

\end{deluxetable}

\begin{deluxetable}{rrrr}
\tablecaption{Best-Fit Transit Times \label{tbl2}}
\tablewidth{0pt}
\tablehead{
\colhead{Transit \#} & \colhead{T$_C$ (HJD)} & \colhead{$\sigma$ (HJD)} & \colhead{Reduced $\chi^2$}}
\startdata
1 &  2453235.49852 & $\pm$0.00068 & 0.83 \\
3 &  2453242.54809 & $\pm$0.00049 & 0.74 \\
4 &  2453246.07367 & $\pm$0.00061 & 0.84 \\
101 &  2453587.97480 & $\pm$0.00074 & 0.91 \\
102 &  2453591.49958 & $\pm$0.00051 & 0.82 \\
103 &  2453595.02365 & $\pm$0.00060 & 0.88 \\
104 &  2453598.54930 & $\pm$0.00053 & 0.79 \\
105 &  2453602.07253 & $\pm$0.00062 & 0.89 \\
106 &  2453605.59805 & $\pm$0.00055 & 0.81 \\
107 &  2453609.12212 & $\pm$0.00056 & 0.91 \\
108 &  2453612.64690 & $\pm$0.00056 & 0.75 \\
109 &  2453616.17165 & $\pm$0.00062 & 0.82 \\
110 &  2453619.69667 & $\pm$0.00065 & 0.79 \\
111 &  2453623.22186 & $\pm$0.00075 & 0.91 \\
112 &  2453626.74617 & $\pm$0.00079 & 1.02 \\

\enddata

\tablecomments{Transit times for 3 transits in 2004 and 12 transits in 2005.  
  Transit 2 was not fully observed due to an onboard software crash, 
  and it has been omitted since an accurate transit time cannot be obtained.}

\end{deluxetable}

\clearpage

\end{document}